\documentclass[12pt]{article}
\usepackage{multirow}
\usepackage[section] {placeins}

\usepackage[papersize={8.5in,11in},hdivide={1in,*,1in},vdivide={1in,*,1in},dvips,mag=1000,nofoot]{geometry}
\usepackage{graphicx}
\usepackage{amsmath,amsfonts,amssymb}
\usepackage{amsthm}
\usepackage{natbib}
\bibpunct{(}{)}{;}{a}{,}{,}

\numberwithin{equation}{section}

\long\def\symbolfootnote[#1]#2{\begingroup
\def\thefootnote{\fnsymbol{footnote}}\footnote[#1]{#2}\endgroup}

\begin{document}

\pagenumbering{gobble}
\clearpage

\noindent \textbf{\large{Bayesian Model-Averaged Benchmark Dose Analysis Via Reparameterized Quantal-Response Models}}

\vspace{8 pt}
\renewcommand\thefootnote{\fnsymbol{footnote}}
\noindent \textbf{Qijun Fang}$^{\textrm{a}}$, \textbf{Walter W.~Piegorsch}$^{\textrm{a,b}}$, \textbf{Susan J.~Simmons}$^{\textrm{c}}$, \textbf{Xiaosong Li}$^{\textrm{c}}$, \textbf{Cuixian Chen}$^{\textrm{c}}$ and \textbf{Yishi Wang}$^{\textrm{c}}$
\\
\noindent $^{\textrm{a}}$\textit{Graduate Interdisciplinary Program in Statistics} and $^{\textrm{b}}$\textit{BIO5 Institute}\\
\textit{University of Arizona, Tucson, AZ, USA}\\
\noindent $^{\textrm{c}}$\textit{Department of Mathematics and Statistics}\\
\textit{University of North Carolina at Wilmington, Wilmington, NC, USA}\\

\vspace{1 pc}

\noindent \textbf{Abstract}\\
\small
\noindent An important objective in biomedical risk assessment is estimation of minimum exposure levels that induce a pre-specified adverse response in a target population. The exposure/dose points in such settings are known as Benchmark Doses (BMDs). Recently, parametric Bayesian estimation for finding BMDs has become popular. A large variety of candidate dose-response models is available for applying these methods, however, leading to questions of model adequacy and uncertainty. Here we enhance the Bayesian estimation technique for BMD analysis by applying Bayesian model averaging to produce point estimates and (lower) credible bounds. We include reparameterizations of traditional dose-response models that allow for more-focused use of elicited prior information when building the Bayesian hierarchy. Performance of the method is evaluated via a short simulation study. An example from carcinogenicity testing illustrates the calculations.

\par

\vspace{2 pc}

 \par

\vspace{1 pc}
\normalsize

\clearpage
\pagenumbering{arabic}
\setcounter{page}{1}

\section{Introduction}\label{sec:intro}
\subsection{Benchmark Risk Analysis}\label{sec:BMD RA}
\indent \indent
An important objective in quantitative risk assessment is characterization of detrimental or adverse responses after exposure to biological, chemical, physical, environmental, or other hazardous agents \citep{ster08}. In this context, the risk is often quantified via a dose-response function, $R(d)$ (also called the \emph{risk function}), which is defined as the probability of exhibiting the adverse effect in a subject, object, or system exposed to a particular dose or exposure level, $d$, of the agent. For biomedical and environmental risk applications, statistical dose-response models are fit to data from bioassays on small mammals or other biological systems, or from epidemiological analyses of human populations at risk.  The observations in such risk assessments are often in the form of proportions. This is the \textit{quantal response setting}, and is our focus here.

A modern approach to quantal-response risk estimation is known as \emph{benchmark analysis}. First introduced for toxicological applications by \cite{crum84}, this method uses a specific functional assumption on $R(d)$ to provide low-dose estimates for the risk.  When applied in biomedical or other public health settings, however, adjustments for spontaneous effects must be introduced, re-expressing the risk in terms of an excess-above-background, exposure-related rate \citep[\S4.2.1]{piba05}.  For instance, with quantal data the \emph{extra risk function} $R_E(d)=\frac{R(d)-R(0)}{1-R(0)}$, where $R(0)$ is the background risk, is frequently employed.  From this, the \emph{benchmark dose} (BMD) is calculated  by inverting $R_E(d)$ at a predetermined level of risk; the latter is called the \emph{benchmark risk} or \emph{benchmark response} (BMR).  In effect, we solve for $d =$ BMD in $R_E$(BMD) = BMR $\in (0,1)$; BMRs between 0.01 and 0.10 are most often seen in practice \citep{epa12}. If the exposure is measured as a concentration, the benchmark point is referred to as a \emph{benchmark concentration} (BMC), if exposure is some sort of quantitative index the benchmark point is a \emph{benchmark index} (BMI), etc.
To illustrate, consider the following example.

\vspace{8pt}
\noindent \textit{Example 1. Benchmarking mammalian carcinogenicity of cumene.}\\
Cumene, the colloquial name for isopropylbenzene (C$_9$H$_{12}$), is a hydrocarbon solvent employed in the production of industrial compounds such as phenol and acetone.  Occupational and industrial exposures to cumene are common, so the U.S. National Toxicology Program (NTP) explored various forms of mammalian toxicity to the chemical \citep{tr542}. For example, Table \ref{tabl:data} displays quantal-response data on induction of lung tumors (alveolar/bronchiolar adenomas and carcinomas) by cumene in laboratory mice after two year chronic inhalation exposure.
%
\linespread{1.2}
\begin{table} [htbp]
\caption{\small Quantal carcinogenicity data:  Alveolar/bronchiolar adenomas and carcinomas in female B6C3F$_1$ mice after inhalation exposure to cumene (C$_9$H$_{12}$).  Source: \cite{tr542}.}\label{tabl:data}
\renewcommand{\arraystretch}{1.25} 
\vspace{8pt} \centering
\begin{small}
\begin{tabular}{l c c c c c}
\hline
Exposure conc.~(ppm), $d_i$  && 0 & 125 & 250 & 500  \\
\hline
Animals with tumors, $Y_i$ & & 4 & 31 & 42 & 46 \\
Animals tested, $N_i$ & & 50 & 50 & 50 & 50 \\
\hline
\end{tabular}
\end{small}
\end{table}

In the table, a clear dose response is evidenced. Of additional interest, however, is calculation of a benchmark exposure level to inform risk characterization of this potential carcinogen.  Notice that the C$_9$H$_{12}$ exposure dose, $d$, is actually a concentration (in ppm) here, and so technically we will compute benchmark concentrations (BMCs) based on the quantal carcinogenicity data. A benchmark analysis of these data appears Section \ref{sec:Example}.

\vspace{8pt}
One critical enhancement in benchmark analysis is construction of statistical confidence (or credible) intervals for the BMD to account for variability in the estimation process.  Driven by public health or other safety considerations, only one-sided, lower limits are employed, denoted as BMDLs \citep{crum95}.  Where needed for clarity, we add a subscript for the BMR level at which each quantity is calculated:  BMD$_{\text{\tiny 100BMR}}$ and BMDL$_{\text{\tiny 100BMR}}$. In this fashion, BMDs and BMDLs are employed for risk characterization by a number of government and private entities.  Their use for quantifying and managing risk with a variety of biological endpoints is growing in both the United States and the European Union \citep{gao01,eu03,oecd06,oecd08}.

\subsection{Parametric Bayesian Benchmark Analysis}\label{sec:MuncMA}

\indent\indent Statistical estimation of the BMD is fairly well-developed under a frequentist schema \citep[\S4.3]{piba05}; however, Bayesian
benchmark analysis has only recently garnered appreciative attention.
Some modern advances can be found in \citet{nauf09,shsm11,shao12,whba12}; and \citet{guha13}. Notably, these sources for calculating Bayesian BMDs generally parameterize the model in terms of standard regression-type quantities.  For example, the logistic dose response $R(d) = (1 + \exp\{-\beta_0 - \beta_1 d\})^{-1}$ is popular, with $\beta_0$ and $\beta_1$ representing the intercept and slope, respectively, of the risk on a logit scale.  Under this sort of traditional parameterization, the Bayesian hierarchy is usually presented with objective and/or improper prior distributions for the unknown $\beta$-parameters. [Some noteworthy exceptions include incorporation at the prior level of historical control information related to $\beta_0$; see \cite{shao12} and \cite{whba12}.]
Indeed, it is unusual for truly informative, risk-analytic, prior knowledge to be available on regression-type parameters such as $\beta_0$ and $\beta_1$, since their interpretation is so generic.  Unfortunately, this strategy can neglect informative prior information available on the true quantity of interest in this setting, the BMD.

Of course, prior information on the BMD may not always exist in practice, and in this case objective priors serve a useful purpose.  When informative prior knowledge is available, however, the Bayesian paradigm can achieve its full potential.
Along these lines, we previously developed reparameterized dose-response models that explicitly incorporate the BMD and other pertinent parameters into the model hierarchy \citep{fapi14}. When applied to Bayesian dose-response modeling, these reparameterizations allow for more practical elicitation of prior information in a benchmark analysis.
Our method assumes, however, that the choice of dose-response function is made without any uncertainty, i.e., that the specification for $R(d)$---reparameterized or otherwise--- is unambiguous and correct.  In practice, the extensive library of quantal dose-response forms for $R(d)$ available to the risk assessor can lead to uncertainty in the model specification.
To mitigate concerns over model adequacy/uncertainty, recent works employ model averaging techniques. Bayesian model averaging \cite[BMA; see][]{homa99} has become popular in this regard, and a number of articles have applied some form of BMA to benchmark analysis; see, e.g., \citet{bano05,moib06,shsm11,shao12,shgi14}; and the references therein.
Most of these employ traditional $\beta$-parameterizations with objective priors, however.  Here, we study the BMA approach for benchmark analysis by using the reparameterized framework from \cite{fapi14}. Section \ref{sec:HierBayes} reviews this reparameterized paradigm for hierarchical Bayesian benchmark analysis. Section \ref{sec:BMA} follows by connecting it with the elements of Bayesian model averaging.
Section \ref{sec:Example} returns to the cumene carcinogenicity data from Table \ref{tabl:data} to explore practical implementation of this Bayesian model-averaged benchmark analysis, while
Section \ref{sec:siml} examines the characteristics of our BMA Bayesian
BMDLs via a short simulation study. Section \ref{sec:disc} ends with a brief discussion.

\section{Hierarchical Bayesian Benchmark Analysis} \label{sec:HierBayes}
\subsection{Bayesian Dose-response Models}\label{sec:BayBen}
\indent\indent Under the quantal-response formulation, denote $Y_i$ as the number of responses, $N_i$ as the number of subjects tested, and $R(d_i)$ as the unknown probability that an individual subject will respond at (ordered) dose $d_i\ge0$, $i=1, \ldots, m$. Assume that, independently, $Y_i\sim \mbox{Bin}(N_i, R(d_i))$. Let $R(d_i)$ depend upon $\boldsymbol{\theta}$, an unknown parameter vector. The joint p.m.f. of $\boldsymbol{Y}=(Y_1, \ldots, Y_m)$ is then:
\begin{equation}\label{eq:jointpmf}
f(\boldsymbol{Y}|\boldsymbol{\theta})=\prod_i^m {N_i \choose Y_i}R(d_i)^{Y_i}(1-R(d_i))^{N_i-Y_i}.
\end{equation}

Previous parametric representations for modeling $R(d)$ in risk-analytic carcinogenicity testing have focused on a suite of eight different functions \citep{whba09b,pian13}. These correspond to popular choices in the U.S.~EPA's BMDS software \citep{davi12}, and are illustrated in Table \ref{tab:t2}.
[Throughout, $\Phi(\cdot)$ represents the c.d.f. from the standard normal distribution.] Notice in the table that models M$_1$--M$_4$ employ only two unknown parameters, while models M$_5$--M$_8$ employ three.
The BMD for each model is obtained by first setting the extra risk function (as defined in Section \ref{sec:intro}) equal to the BMR and then solving for $d$.

\begin{table}[!ht]
\begin{center}
\caption{\small Common quantal-response models in toxicological and carcinogenic risk assessment. Note: BMR$\in (0,1)$ is the benchmark response and BMD is the benchmark dose} \vspace{8pt}
\label{tab:t2}
\scalebox{0.8}{
\begin{tabular}{clccc}
\hline
\textbf{Code} &\textbf{Name} &\textbf{Risk Function,} {\boldmath $R(d)$}&\textbf{BMD} &\textbf{Constraints} \\
\hline
M$_1$ & Logistic & $(1+\exp\{-\beta_0-\beta_1d\})^{-1}$ & $\frac{1}{\beta_1}\ln\left(\frac{1\,+\,\text{\scriptsize BMR} e^{-\beta_0}}{1\,-\,\text{\scriptsize BMR}}\right)$ & ---\\
\\
M$_2$ & Probit & $\Phi(\beta_0+\beta_1d)$ & $\frac{\Phi^{-1}\{[1-\Phi(\beta_0)]\text{\scriptsize BMR}+\Phi(\beta_0)\}-\beta_0}{\beta_1}$ & ---\\
\\
M$_3$ & Quantal-Linear & $1-\exp\{-\beta_0-\beta_1d\}$ & $-\frac{1}{\beta_1}\ln(1-\text{\footnotesize BMR})$ & $\beta_0\ge0$, $\beta_1\ge0$\\
\\
M$_4$ & Quantal-Quadratic & $\gamma_0+(1\!-\!\gamma_0)(1-e^{-\beta_1d^2})$ & $\sqrt{-\frac{1}{\beta_1}\ln(1-\text{\footnotesize BMR})}$ & $0\le\gamma_0<1$, $\beta_1\ge0$\\
\\
M$_5$ & Two-Stage & $1-\exp\{-\beta_0-\beta_1d-\beta_2d^2\}$ & $\frac{-\beta_1+\sqrt{\beta_1^2-4\beta_2\ln(1-\text{\scriptsize BMR})}}{2\beta_2}$ & $\beta_0,\beta_1,\beta_2\ge0$\\
\\
M$_6$ & Log-Logistic & $\gamma_0+\frac{1-\gamma_0}{1\,+\,\exp(-\beta_0-\beta_1\ln\{d\})}$ & $\exp\left\{\frac{\ln\left(\frac{\text{\tiny BMR}}{1-\text{\tiny BMR}}\right)-\beta_0}{\beta_1}\right\}$ & $0\le\gamma_0<1$, $\beta_1\ge0$\\
\\
M$_7$ & Log-Probit & $\gamma_0+(1\!-\!\gamma_0)\Phi(\beta_0+\beta_1\ln\{d\})$ & $\exp\left\{\frac{\Phi^{-1}(\text{\scriptsize BMR})-\beta_0}{\beta_1}\right\}$ & $0\le\gamma_0<1$, $\beta_1\ge0$\\
\\
M$_8$ & Weibull & $\gamma_0+(1\!-\!\gamma_0)(1-\exp\{-e^{\beta_0}d^{\beta_1}\})$ & $\exp\left\{\frac{\ln(-\ln[1-\text{\scriptsize BMR}])-\beta_0}{\beta_1}\right\}$ & $0\le\gamma_0<1$, $\beta_1\ge1$\\
\hline \vspace{-36pt}
\end{tabular}
}
\end{center}
\end{table}
In applications of Bayesian benchmark analysis to quantal data,
objective forms for the prior p.d.f.~$\pi(\boldsymbol{\theta})$ are common, as indicated earlier. These typically appear as diffuse Gaussian priors on the $\beta$-parameters.
From this, the joint posterior distribution for $\boldsymbol{\theta}$ is obtained from Bayes formula \citep[\S7.2.3]{cabe02}.
An advantage here is that objective priors are usually easy to apply: although they generally lead to intractable integrals,
computer-intensive operations such as Markov chain Monte Carlo (McMC) methods can produce a corresponding sample from the joint posterior of $\boldsymbol{\theta}$ \citep{roca11}.  If the sample is sufficiently large and stable, the output can be used to approximate the posterior, from which inferences on the BMD may be conducted.

As we suggest above, a disadvantage with the traditional parameterizations in Table \ref{tab:t2} is that the $\beta$-parameters may have unclear subject-matter interpretations if those parameters are not the target quantities of interest. If informative prior information were available on the risk-analytic quantities under study, the ambiguous interpretation(s) of these traditional, regression-type parameterizations makes incorporation of such information more difficult.  This may hinder effective application of the Bayesian approach in this benchmark setting.

\subsection{Reparameterizing the Quantal Response Models}\label{sec:Reparameterization}
\indent\indent For benchmark risk analysis, we argue \citep{fapi14} that substantive prior knowledge may be available more often than usually considered, but not in the form of information on regression-type $\beta$-parameters. Instead, a risk assessor, toxicologist, or other domain expert would typically have prior knowledge about the target parameter, the BMD, and possibly also about other application-specific values.
Our goal is to utilize the domain expert's prior knowledge for making inferences on the BMD.  To do so, we follow \cite{fapi14} and reparameterize $R(d)$ in terms of ostensibly meaningful parameters whose prior distributions are more intuitive to elicit in practice.

For the dose-response models with two parameters in Table \ref{tab:t2}, we reparameterize in terms of the target value, BMD (denoted in the sequel as $\xi$), and the background risk, say, $\gamma_0 = R(0)$.
Thus $\boldsymbol{\theta}$ becomes the vector $[\xi~ \gamma_0]{}^\text{\scriptsize T}$.

For the models with three unknown parameters in Table \ref{tab:t2}, we reparameterize with $\xi$ = BMD, $\gamma_0 = R(0)$, and a parameter $\gamma_1$ defined as $R(d_\ell)$ for some non-zero dose level $d_\ell$.
Unless otherwise specified, we set $d_\ell$ to the highest dose, so $\gamma_1 = R(d_m)$. Thus, we take $\boldsymbol{\theta} = [\xi~\gamma_0\;\gamma_1]{}^\text{\scriptsize T}$.  The latter two quantities are technically nuisance parameters as far as the BMD is concerned, but one or both are nonetheless likely to be associated with non-trivial prior information; e.g., historical control data may inform $\gamma_0 = R(0)$ \citep{whba12,shao12}.

For instance, consider a highly popular three-parameter model from carcinogenicity assessment, the two-stage version of the multi-stage model \citep{ardo54, nitc07}.  This is model M$_5$ in Table \ref{tab:t2}:
\begin{equation} \label{eq:M5orig}
R(d) = 1 - \exp\{-\beta_0 - \beta_1 d - \beta_2 d^2\},
\end{equation}
where $\beta_j \ge 0 \text{ for all } j=0,1,2$.  It is fairly straightforward to apply our reparameterization strategy and represent the three $\beta$-parameters in terms of $\boldsymbol{\theta} = [\xi~\gamma_0\;\gamma_1]{}^\text{\scriptsize T}$; we explicate this in a Supplementary Document.  The result is
\begin{equation} \label{eq:M5}
R(d)=\gamma_0+(1-\gamma_0)\left[1-\exp\left\{\frac{C_5d_m d (d_m-d)+\Gamma_5 \xi d (\xi-d)}{\xi d_m(\xi-d_m)}\right\}\right]
\end{equation}
where $\Gamma_5=\log\left(\frac{1-\gamma_1}{1-\gamma_0}\right)$ and $C_5=-\log(1-\mbox{BMR})$.  Admittedly, this is much less compact than the common form in \eqref{eq:M5orig}, but it nonetheless gives an expression that can be manipulated effectively for Bayesian benchmark analysis.

In similar fashion, the remaining models in Table \ref{tab:t2} can be reparameterized as follows:
\begin{small}
\begin{description}
\item[M$_1$:] $R(d) = \left[1+\exp\left\{-\mbox{logit}(\gamma_0) - \frac{d}{\xi}\log\left(\frac{1+\exp\{-\mbox{\scriptsize logit}(\gamma_0)\}\cdot \mbox{\scriptsize BMR}}{1-\mbox{\scriptsize BMR}}\right)\right\}\right]^{-1}$.

\item[M$_2$:] $R(d)=\Phi\left(\Phi^{-1}\{\gamma_0\}+\frac{\Phi^{-1}\{\mbox{\scriptsize BMR}[1-\gamma_0]+\gamma_0\}-\Phi^{-1}(\gamma_0)}{\xi}d\right).$

\item[M$_3$:] $R(d)=1-\exp\left\{\log(1-\gamma_0)+\frac{\log(1-\mbox{\scriptsize BMR})}{\xi}d\right\}.$

\item[M$_4$:] $R(d) =\gamma_0+(1-\gamma_0)\left(1-\exp\left\{\frac{\log(1-\mbox{\scriptsize BMR})}{\xi^2}d^2\right\}\right).$

\item[M$_5$:] See Equation \eqref{eq:M5}.

\item[M$_6$:] $R(d)=\gamma_0+(1-\gamma_0)\left[1+\exp\left\{\frac{C_6[\log d_m-\log d]+\Gamma_6[\log\xi-\log d]}{\log\xi-\log d_m}\right\}\right]$,\\
where $\Gamma_6=\log\left(\frac{1-\gamma_1}{1-\gamma_0}\right)$ and $C_6=\log\left(\frac{\small \mbox{\scriptsize BMR}}{\small 1-\mbox{\scriptsize BMR}}\right)$.

\item[M$_7$:] $R(d) = \gamma_0+(1-\gamma_0)\Phi\left(\frac{C_7[\log d_m-\log d]+\Gamma_7[\log d-\log\xi]}{\log d_m-\log\xi}\right)$,\\
where $\Gamma_7=\Phi^{-1}\left(\frac{\gamma_1-\gamma_0}{1-\gamma_0}\right)$ and $C_7=\Phi^{-1}(\mbox{BMR})$.

\item[M$_8$:] $R(d) = \gamma_0+(1-\gamma_0)\left(1-\exp\left\{-\exp\left(\frac{C_8[\log d_m-\log d]+\Gamma_8[\log d-\log\xi]}{\log d_m-\log\xi}\right)\right\}\right)$,\\
where $\Gamma_8=\log\left(-\log\left(\frac{1-\gamma_1}{1-\gamma_0}\right)\right)$ and $C_8=\log(-\log(1-\mbox{BMR}))$.
\end{description}
\end{small}
Here again, these reparameterizations present more burdensome notation for $R(d)$.  The explicit incorporation of the target parameter $\xi$ and well-understood quantities such as $\gamma_0$ and $\gamma_1$ allows us, however, to formulate a clearer, more application-oriented hierarchical model, from which to produce inferences on $\xi$.

\subsection{Bayesian Benchmark Analysis under Reparameterized Quantal Response Models}\label{sec:BayRepar}

\indent\indent
Generically, under our reformulation we assign a joint p.d.f.~to $\boldsymbol{\theta}$: $\pi(\boldsymbol{\theta}) = \pi(\xi, \gamma_0)$ for the two-parameter quantal-response models, or $\pi(\boldsymbol{\theta}) = \pi(\xi, \gamma_0, \gamma_1)$ for the three-parameter models.  Mimicking previous Bayesian constructions for benchmark analysis \citep{shsm11,shao12}, we assume the unknown parameters enter into the prior independently, so that $\pi(\xi, \gamma_0) = \pi(\xi)\pi(\gamma_0)$ or $\pi(\xi, \gamma_0, \gamma_1)=\pi(\xi)\pi(\gamma_0)\pi(\gamma_1)$.

For the non-negative, target quantity $\xi$ we employ an inverse gamma prior: $\xi \sim \mbox{\textit{IG}}(\alpha, \beta)$ with marginal prior density $\pi(\xi|\alpha,\beta) = \frac{\beta^\alpha}{\Gamma(\alpha)}\xi^{-(\alpha+1)}e^{-\beta/\xi}I_{(0,\infty)}(\xi)$, where $\Gamma(a)$ is the usual Gamma function and $I_{\mathbb{A}}(x)$ is the indicator function that returns 1 if $x \in \mathbb{A}$ and 0 otherwise.  For the probability parameter $\gamma_0 = R(0)$ we take $\gamma_0 \sim \mbox{\textit{Beta}}(\psi, \omega)$ with marginal prior $\pi(\gamma_0|\psi,\omega) = \frac{\Gamma(\psi+\omega)}{\Gamma(\psi)\Gamma(\omega)}\gamma_0^{\psi-1}(1-\gamma_0)^{\omega-1}I_{(0,1)}(\gamma_0)$. Where needed (models M$_5$--M$_8$) we similarly set $\gamma_1\sim Beta(\kappa, \lambda)$.

The various hyperparameters---$\alpha$, $\beta$, $\psi$, $\omega$, $\kappa$ and $\lambda$---require complete specification for implementation of the model as we propose it. To do so, we first attempt to elicit each prior by incorporating subject-matter knowledge of the associated quantities from either an individual domain expert or previous information in the literature (or both). Since the IG and beta priors each have two parameters, two associated quantities are needed. Based on interactions with toxicologists and risk assessors, we find that for the target parameter $\xi$, specification of the first/lower quartile ($Q_1$) along with the median ($Q_2$) of the IG prior is most propitious \citep{fapi14} . Similarly, for the Beta prior on $\gamma_0$ (and $\gamma_1$) we also elicit the two quartiles $Q_1$ and $Q_2$. We then numerically solve the resulting system of equations for the pertinent hyper-parameters given the two quartiles. Details are provided in the Supplementary Document.

When elicitation is not possible, we default to objective specifications for the prior densities. We favor a simple approach: for an objective prior on $\xi$, use the proper prior, $\xi \sim IG(0.001,0.001)$.
This IG prior is a popular suggestion in the literature for right-skewed, positive values \citep[\S1.2]{lamb05,chjo11}, such as the BMD. It is in effect an approximation of the conventional improper reciprocal prior for positive quantities, i.e.,  $\pi(\xi)\propto 1/\xi$ \citep[\S9.17]{ohag94}.
To ensure computational stability, following guidance from \citet{fapi14}, we perform all our hierarchical calculations with doses scaled so that the maximum administered dose equals 1.
For an objective prior on either $\gamma_0$ and/or $\gamma_1$, we choose another conventional, objective prior for proportions, the univariate Jeffreys prior:  $ \mbox{\textit{Beta}}\bigl(\frac12,\frac12\bigr)$.

From these, the posterior p.d.f.~for ${\boldsymbol \theta}$ under the two-parameter dose-response models is
\begin{equation}\label{eq:post2}
\pi(\xi, \gamma_0|{\boldsymbol Y})=\frac{\prod_{i=1}^n{N_i \choose Y_i}{R(d_i)}^{Y_i}{[1-R(d_i)]}^{N_i-Y_i}}{m(\boldsymbol Y)}
\frac{\beta^\alpha e^{-\beta/\xi}}{\Gamma(\alpha)\xi^{\alpha+1}}
\frac{\Gamma(\psi+\omega)}{\Gamma(\psi)\Gamma(\omega)}\gamma_0^{\psi-1}(1-\gamma_0)^{\omega-1}
\end{equation}
over $\xi > 0$ and $0<\gamma_0 <1$.  A similar form emerges when considering the three-parameter dose-response models.

The denominator of $\pi(\xi, \gamma_0|{\boldsymbol Y})$ in \eqref{eq:post2} contains the marginal likelihood
\[m(\boldsymbol Y) = \int_0^\infty\int_0^1 f({\boldsymbol Y}|\xi, \gamma_0)\pi(\xi|\alpha,\beta)\pi(\gamma_0|\psi,\omega)d\gamma_0d\xi,\]
where $f({\boldsymbol Y}|\xi, \gamma_0)$ is the binomial likelihood.  This marginal is intractable under our elicited prior structure, unfortunately, and to evaluate \eqref{eq:post2} we turn to Monte Carlo posterior approximations that produce simulated realizations ${\boldsymbol \theta}_k, \,k=1,\ldots,K$ of the parameter vector from the posterior distribution.  Our choice for the Monte Carlo technique employs an adaptive Metropolis (AM) strategy, which tunes the variance of the underlying Metropolis proposal density adaptively when generating ongoing draws of the chain.  From our experience with a variety of such methods, we favor a global AM procedure with componentwise adaptive scaling described by \citet{anth08}.  Details are similar to those in \cite{fapi14} and are given in the Supplementary Document.


We monitor convergence of the AM sample to the posterior joint distribution via standard MC diagnostics. We include a burn-in over the first $K_0-1$ draws, where we choose $K_0$ to be much smaller than $K$. $K_0$ is determined
from a series of diagnostics given by \citet{gewe92}. (Again, details can be found in the Supplementary Document.)  If the diagnostics indicate that convergence is not evidenced, we flag the result as an `algorithm failure.'  In \citet{fapi14} we found such failures to be rare events; they usually appear with very shallow dose-response patterns. Our experience also suggests that $K = 100,000$ iterations of the AM chain, including burn-in, generally provide stable results.  From these, we use the remaining $K^* = K - K_0 +1$ draws to approximate the joint posterior for ${\boldsymbol \theta}$.

For estimating the BMD, one can follow a standard approach and select the Bayes estimator as the posterior mean of the AM sample \citep[\S7.2.3]{cabe02}.  Other formulations are also possible; see \citet{fapi14}.
For the corresponding Bayesian BMDL, say, \underline{$\xi$}$_{\text{\tiny 100BMR}}$, we find the one-sided, lower, $100(1-\alpha)$\% credible limit on $\xi$, satisfying $P(\xi >$ \underline{$\xi$}$_{\text{\tiny 100BMR}}|\boldsymbol Y)=1 - \alpha$. At the traditional level of $\alpha = 0.05$, we desire the lower 5th percentile of $\pi(\xi|{\boldsymbol Y})$, and we approximate this using the lower 5th percentile from our Monte Carlo sample of $\xi$.  If $\{\xi_{(k)}\}_{k=1}^{K^*}$ denotes the ordered values from the retained chain, our Bayesian BMDL takes the form \underline{$\xi$}$_{\text{\tiny 100BMR}} = \xi_{(\lfloor 0.05K^* \rfloor)}$, where $\lfloor x \rfloor$ is the floor function that returns the largest integer smaller than $x$.

\section{Bayesian Model Averaging}\label{sec:BMA}
\indent\indent The selection of parametric forms for $R(d)$ presented in \S\ref{sec:Reparameterization} illustrates the wide variety of dose-response functions available to the risk analyst.  Many of these operate well at (higher) doses near the range of the observed quantal outcomes; however, they can produce wildly different estimates on BMDs at very small levels of risk \citep{faus97,kako00}. The corresponding issue of model uncertainty has bedevilled benchmark analysts since its introduction in the mid-1980s.  Some users have turned to formal model selection procedures to derive the BMD \citep{davi12}. This is a natural option, although it relies on a reliable selection criterion. In fact, model selection based on the popular Akaike information criteria (AIC) \citep{akai73} has been shown to select incorrect models for BMD estimation almost as often as it selects correct models \citep{wepi12}. As an alternative to simple model selection methods, and to provide a more model-robust option for BMD estimation, we consider here a (Bayesian) model averaging approach.

Suppose $Q>1$ quantal-response models are under consideration such as the $Q = 8$ models in Table \ref{tab:t2}. These form an uncertainty class $U_Q$, with individual model elements $M_q\; (q = 1,\ldots,Q)$.  Following \cite{homa99}, the model-averaged posterior density for $\xi$ is defined as a mixture of marginal posterior densities for $\xi$ under each model, $f(\xi|\boldsymbol{Y}, M_q)$, with the individual-model posterior probabilities $P(M_q|\boldsymbol{Y})$ employed as weights:
\begin{equation}\label{eq:BMA}
f(\xi|\boldsymbol{Y}, U_Q)=\sum_{q=1}^Qf(\xi|\boldsymbol{Y}, M_q)P(M_q|\boldsymbol{Y}).
\end{equation}
The posterior model probabilities in \eqref{eq:BMA} are obtained through Bayes' formula:
\begin{equation}\label{eq:PMP}
P(M_q|\boldsymbol{Y})=\frac{m(\boldsymbol{Y}|M_q)P(M_q)}{\sum_{q=1}^Q m(\boldsymbol{Y}|M_q)P(M_q)},
\end{equation}
where $P(M_q)$ are the prior model probabilities and $m(\boldsymbol{Y}|M_q)$ are the marginal likelihoods for each $q$th model. If the risk assessor has no preference for any particular model,
a reasonable default views the $M_q$s as equally valid, so that $P(M_q)=1/Q \quad\forall q$. The posterior model probabilities are then simply $P(M_q|\boldsymbol{Y}) = \frac{m(\boldsymbol{Y}|M_q)}{\sum_{q=1}^Q m(\boldsymbol{Y}|M_q)}.$

Clearly, the challenge with this sort of Bayesian model averaging (BMA) is to estimate the $P(M_q|\boldsymbol{Y})$ values. This requires accurate estimation of the marginal likelihood $m(\boldsymbol{Y}|M_q)$.  Unfortunately, under our reparameterized hierarchy the associated integral is intractable. To approximate it, we employ the geometric bridge sampler of \citet{meng96}; also see \citet{lope04}. Bridge sampling is relatively convenient and quite suitable for estimating marginal likelihoods from an AM sample. (Greater detail is provided in the Supplementary Document.)
As pointed out by \citet{homa99}, under a squared error loss function the BMA estimate of $\xi$ is then straightforward to calculate.  Simply take the weighted average of the individual-model, 
posterior sample mean estimates of $\xi_{\text{\tiny 100BMR}}$ across the uncertainty class:
\begin{equation} \label{eq:BMABMD}
\bar{\xi}_{\text{\tiny 100BMR}} = \sum_{q=1}^Qw_q \hat{\xi}_{\text{\tiny 100BMR};q} \, ,
\end{equation}
where the weights are the posterior model probabilities, $w_q = P(M_q|\boldsymbol{Y})$.  The primary focus in benchmark analysis is, however, the BMDL.
To find a Bayesian BMA-based 95\% BMDL we write $P(\xi\le \mbox{\underline{$\xi$}}_{\mbox{\tiny 100BMR}}|\boldsymbol{Y}, U_Q) = 0.05,$
where $\xi$ has p.d.f.~defined by \eqref{eq:BMA}.  From our AM sample, we approximate this via
\[P(\xi\le \mbox{\underline{$\xi$}}_{\mbox{\tiny 100BMR}}|\boldsymbol{Y}, U_Q)\approx\sum_{q=1}^Q w_q\frac{1}{K^*_q}\sum_{j=1}^{K^*_q}I\{\xi_j\le\mbox{\underline{$\xi$}}_{\mbox{\tiny 100BMR}}|M_q\},\]
where $K_q^*$ is the size of the AM sample from model $M_q$ (after burn-in) and $I$ denotes the indicator function.
Thus the estimated BMA BMDL satisfies the equation
\[\sum_{q=1}^Q w_q\frac{1}{K^*_q}\sum_{j=1}^{K^*_q}I\{\xi_j\le\mbox{\underline{$\xi$}}_{\mbox{\tiny 100BMR}}|M_q\}=0.05 .\]
The solution is found numerically.
(Again, technical details are given in the Supplementary Document.)

\section{Example: Benchmarking Carcinogenicity of Cumene}\label{sec:Example}

To illustrate our hierarchical BMA approach with the $Q=8$ reparameterized models in \S\ref{sec:Reparameterization}, we returned to the cumene carcinogenicity data in Table \ref{tabl:data}.  The BMR was set to the standard default level of 0.10 \citep{epa12},
and for the computations all experimental doses were scaled such that the maximum dose was equal to 1.
With input from domain experts, we based prior elicitation for $\xi$ and $\gamma_0$ on existing background in the toxicological literature.  (Specifics are given in the Supplemental Document.)  This led to the prior distributions $\xi\sim IG(0.53,0.13)$ and $\gamma_0\sim Beta(1.36,12.31)$. For the three-parameter models, we required an additional prior specification for $\gamma_1 = R(d_m)$
at $d_m = 1$ (i.e., $d =500$ ppm on the original scale).
Unfortunately, no prior information was available on potential response of the tested animals at this (or any listed) dose of cumene \citep{tr542}.  Thus we defaulted to use of an objective prior:  $\gamma_1\sim Beta(\frac12,\frac12)$.

With this prior structure in place, we generated a Monte Carlo AM sample for each of the eight models.
We encountered no algorithm failures, and the eight generated chains all passed our convergence diagnostic tests.  This produced burn-ins of $10,000$ initial iterations for each chain, allowing us to operate with $90,000$ AM draws for each model. Using the methods described above, this led to model-specific, posterior-mean benchmark concentrations (BMCs) as reported in Table \ref{tab:Exampleres}
(all final BMC values are rescaled to the original dose metric).
The table also lists the corresponding 95\% model-specific BMCLs, along with the posterior model probabilities/weights, $w_q = P(M_q|\boldsymbol{Y})$, assuming uniform prior model probabilities for each model.  The eight BMC were then used to calculate the BMA BMC $\bar{\xi}_{10} = 27.4074$ ppm by using \eqref{eq:BMABMD}.
The corresponding 95\% BMCL is \underline{$\xi$}$_{10} = 15.1927$ ppm. These two values are also reported at the bottom of Table \ref{tab:Exampleres}.
\begin{table}[ht]
\caption{\small BMC estimates based on posterior means and 95\% BMCLs (in ppm) from each reparameterized model in \S\ref{sec:Reparameterization}, along with corresponding Bayesian model averaged (BMA) BMDL, for cumene carcinogenesis example.  The BMR is set to 0.10.}\label{tab:Exampleres}
\vspace{10pt} \centering
\begin{small}
\begin{tabular}{cccc}
  \hline
 Model &  BMC$_{10}$ & BMCL$_{10}$ & $P(M_q|\boldsymbol{Y})$ \\
  \hline
   M$_1$ & 43.2752 & 35.5991 & 0.00044 \\
   M$_2$ & 44.7192 & 37.6845 & 0.00005 \\
   M$_3$ & 18.0881 & 14.7567 & 0.22887 \\
   M$_4$ & 76.3691 & 66.2304 & 0.00000 \\
   M$_5$ & 21.2154 & 16.2568 & 0.01356 \\
   M$_6$ & 31.1642 & 15.9229 & 0.36905 \\
   M$_7$ & 30.1092 & 15.3244 & 0.34956 \\
   M$_8$ & 24.2385 & 17.0606 & 0.03846\vspace{1pt}\\
  BMA   & 27.4074 & 15.1927 &  \\
  \hline \vspace{-12pt}
\end{tabular}
\end{small}
\end{table}

From the table, we see that 
the posterior model probabilities vary widely with these data, suggesting that certain models may provide better-quality estimates than others.  The three-parameter log-logistic and log-probit give the highest weights, both near 35\%, followed by the two-parameter quantal linear model at 23\%.  All other models show posterior probabilities below 5\%.
The BMA 95\% BMCL is 15.1927 ppm, lying within the range of the individual model BMCLs and closer to the individual lower limits associated with higher-weight models.

The ramifications with these data for the risk analyst are substantial: had a choice of the poorly-fitting logistic, probit, or quantal-quadratic models been made for analyzing these data, the consequent BMC and BMCL would have been far too large.  By integrating information across the various models, however, a more-tempered, model-robust estimate is produced from which further risk analytic calculations on cumene carcinogenicity can be conducted.  Corroborating reports by many who have come before, we find that BMA adjustment frees the risk assessor from the selection biases, model inadequacies, and inferential uncertainties one encounters when committing to only a single parametric model to perform the benchmark analysis.

\section{Performance Evaluations}\label{sec:siml}
\subsection{Simulation design}\label{sec:siml design}

\indent\indent To explore the features of our Bayesian model-averaged BMD/BMDLs in further detail, we conducted a short simulation study.
We fixed the BMR at the standard level of BMR = 0.10 \citep{epa12} and operated at a credible level of 95\%. The doses were set to four levels: $d_1$ = 0, $d_2$ = 0.25, $d_3$ = 0.5, $d_4$ = 1, corresponding to a standard design in cancer risk experimentation \citep{port94}.  Equal numbers of subjects, $N_i = N$, were taken at each dose group.  We considered three different per-dose sample sizes:  $N$ = 25, 50, or 1000; the latter approximates a `large-sample' setting, while the former two are more commonly seen with toxicological investigations such as that in the cumene carcinogenicity example. As throughout, all of our calculations were performed within the \verb|R| programming environment \citep{r12}.

For the true dose-response model $R(d)$, we used each of the eight quantal-response functions in Table \ref{tab:t2}.
We considered two different dose-response patterns, from a larger collection studied by \citet{wepi12}.  The first pattern (P-I) was a moderately increasing response with $\gamma_0 = R(0) = 0.05$, $R(\frac12) = 0.30$, and $R(1) = 0.50$.  The second pattern (P-II) was more-broadly increasing, with $\gamma_0 = R(0) = 0.10$, $R(\frac12) = 0.50$, and $R(1) = 0.90$.
The
resulting parameter configurations for the various models are given in Table \ref{SimConfig}.
\begin{table}[!htb]
\caption{\small Models and configurations for simulation study. Model codes are from Table \ref{tab:t2}} \label{SimConfig}
\vspace{8pt} \centering
\begin{footnotesize}
\begin{tabular}{cccccccccc}
\hline
 & & \multicolumn{8}{c}{Dose-Response Model} \vspace{1pt}\\
Configuration & Parameters & M$_1$ & M$_2$ & M$_3$ & M$_4$ & M$_5$ & M$_6$ & M$_7$ & M$_8$\\
\hline
\multirow{3}{*} {P-I} & $\gamma_0$ & 0.05  & 0.05  & 0.05  & 0.05  & 0.05  & 0.05  & 0.05  & 0.05  \\
                      & $\xi$      & 0.3974& 0.3567& 0.1642& 0.4052& 0.1783& 0.2083& 0.2267& 0.1852\\
                      & $\gamma_1$ &  ---   &  ---   &  ---   &  ---   & 0.50  & 0.50  & 0.50  & 0.50\\
\hline
\multirow{3}{*} {P-II}& $\gamma_0$ & 0.10  & 0.10  & 0.10  & 0.10  & 0.10  & 0.10  & 0.10  & 0.10  \\
                      & $\xi$      & 0.1700& 0.1575& 0.0480& 0.2190& 0.1925& 0.2760& 0.2794& 0.2025\\
                      & $\gamma_1$ &  ---   &  ---   &  ---   &  ---   & 0.90  & 0.90  & 0.90  & 0.90\\
\hline
\end{tabular}
\end{footnotesize}
\end{table}

At each of the 8 (models) $\times$ 2 (configurations) $\times$ 3 (sample sizes) = 48 combinations, we simulated 2000 pseudo-binomial quantal-response data sets via \verb|R|'s \verb|rbinom| function. Then for each data set, we generated eight AM samples, one from each model $M_q$.  We then calculated the
model-specific 95\% BMDL \underline{$\xi$}$_{\mbox{\tiny 10};q}$ as the lower $5^{\mbox{\scriptsize th}}$ percentile of each model's AM sample. For simplicity, no prior elicitation was applied in the simulations and so
we employed only objective priors for $\xi$,  $\gamma_0$ and $\gamma_1$, i.e., $\xi\sim IG(0.001,0.001)$, $\gamma_0\sim Beta(\frac12,\frac12)$, and $\gamma_1\sim Beta(\frac12,\frac12)$, to represent parameter uncertainty.

Assuming uniform prior model probabilities, the posterior model probabilities were calculated using geometric bridge sampling. We employed these posterior probabilities $P(M_q|\boldsymbol{Y})$ as weights 
and computed the BMA 
BMDLs according to the
methods described in \S\ref{sec:BMA}.

\subsection{Simulation Results}\label{sec:MC res}

\begin{figure}[!htbp]
\begin{center}
\includegraphics[scale=0.35]{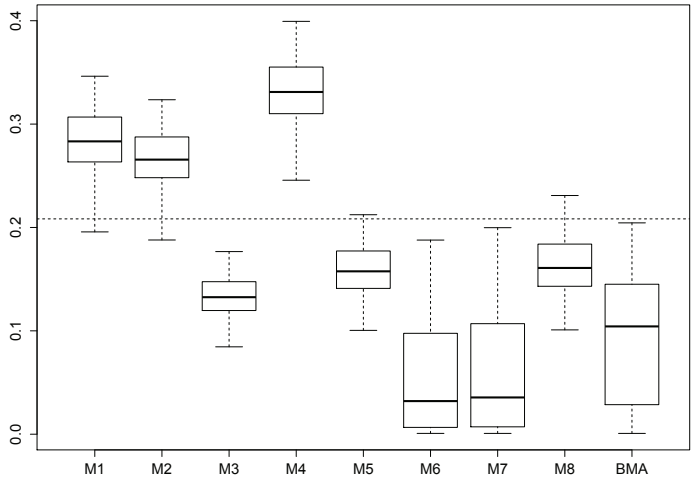}
\end{center}
\caption{\small Modified Box plots for 95\% individual-model BMDLs and BMA BMDL using simulated data from model M$_6$, configuration P-I, and sample size $N = 50$. (See text for details.) BMR is set to 0.10.  Dashed horizontal line indicates target BMD$_{10}$ under this model configuration.}\label{fig:BoxM6D50}
\end{figure}

\indent\indent
We studied how the BMA BMDL compared to the corresponding generating value of $\xi_{10}$ in Table \ref{SimConfig}.
In some sense,
we expect 95\% of these lower credible limits to lie below $\xi_{10}$ and using our simulations as a guide, we queried how often this occurred.  Figure \ref{fig:BoxM6D50} provides representative boxplots of the 2000 simulated BMA BMDLs under model M$_6$ (log-logistic) and configuration P-I at the popular per-dose sample size of $N = 50$.  Mimicking a device employed by \citet{wepi12}, the boxplots are asymmetrically modified so that their upper whiskers stop at the 95th percentile of the 2000 simulated BMA BMDLs.
(The lower whiskers rest at the minimum BMDL.  The hinges and median bar are the usual quartiles.)  Thus the `goal' for each boxplot is to locate its upper whisker as close to, but not greatly exceeding, the generating value of $\xi_{10}$.
A horizontal dashed line in the figure marks this $\xi_{10}$ target.
%

The eight modified boxplots in Figure \ref{fig:BoxM6D50} correspond to model-specific 95\% BMDLs calculated under each of the eight models in Table \ref{tab:t2}.  The modified boxplot at far right gives the result for our 95\% BMA BMDL.  The graphic illustrates the consequences and ambiguities of single-model uncertainty when calculating Bayesian BMDLs.
Recall that M$_6$ is the generating model
and as anticipated, the M$_6$ boxplot for our single-model Bayesian BMDL displays acceptable characteristics: its modified upper whisker is slightly below the $\xi_{10}$ target.  It also exhibits a right skew, however, with median BMDL conservatively lower than any of the others in the figure.

The figure also illustrates the perplexing operating characteristics of single-model BMDLs under model misspecification: when M$_1$, M$_2$, and (particulary) M$_4$ are employed individually in the hierarchy, their BMDLs badly overestimate the $\xi_{10}$ target.  BMDLs using models M$_5$ and M$_8$ are somewhat more-stable; BMDLs using model M$_3$ show the least variation while locating conservatively below the $\xi_{10}$ target.  Analogous instances of stable/unstable single-model BMDLs occurred for all the model-configuration combinations we studied, but with no clear pattern as to which models operated well or poorly when misspecified. [Complete results from all 48 configurations are available in \cite{fang14}.]

On balance, however, the modified boxplot for the BMA BMDL at the far right of Figure \ref{fig:BoxM6D50} displays reasonable operating characteristics.  As desired, its modified upper whisker lies just below the target $\xi_{10}$, and it locates a broader collection of \underline{$\xi$}$_{10}$ limits closer to that target, without exceeding it, than even the correct-model M$_6$ boxplot.  (Admittedly, the M$_3$ and perhaps M$_5$ boxplots display even better performance here; however, these models did not exhibit consistently enhanced performance across all the configurations we studied.)

We find that unequivocal commitment to a specific dose-response model when there is any possibility of it being misspecified can lead, as often as not, to substantial overestimation of the benchmark point(s).
Misspecifying the model sometimes produces acceptable lower limits, but we were not able to identify any predictable pattern of such among our simulation results.  [This corroborates similar indications by \citet{wepi12} for single-model, frequentist BMDL calculations.]  By contrast, the Bayesian BMA BMDL provides a reasonable compromise.

\begin{figure}[!htbp]
\begin{center}
\includegraphics[scale=0.35]{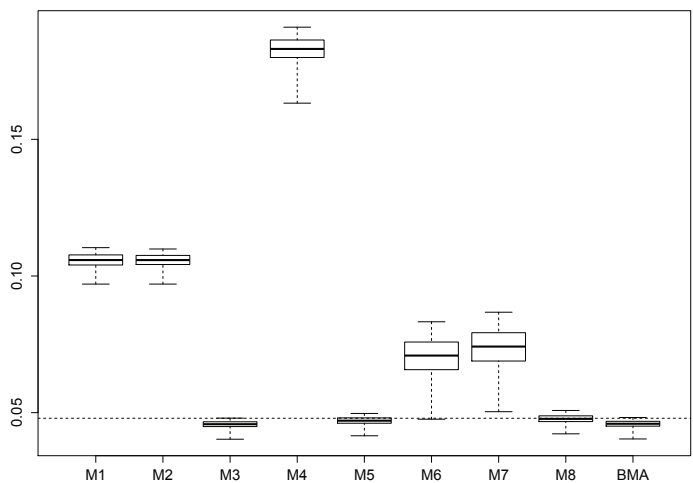}
\end{center}
\caption{\small Modified Box plots for 95\% individual-model BMDLs and BMA BMDL using simulated data from model M$_3$, configuration P-II, and sample size $N = 1000$. (See text for details.) BMR is set to 0.10.  Dashed horizontal line indicates target BMD$_{10}$ under this model configuration.}\label{fig:BoxM3F1000}
\end{figure}

Figure \ref{fig:BoxM3F1000} extends this analysis to the large-sample setting, displaying a similar graphic when the per-dose sample size is $N = 1000$.  Here, we highlight results for model M$_3$ under configuration P-II.  As we might expect with such a large $N$, no single-model fit exhibits acceptable performance, save that of the correct model (M$_3$) and possibly models M$_5$ and M$_8$.  (Notice that model M$_3$ is a special case of both M$_5$, with $\beta_2 = 0$, and M$_8$, with $\beta_1 = 1$, in Table \ref{tab:t2}.  As in Figure \ref{fig:BoxM6D50}, these three models sometimes perform similarly, although we also found cases where their operating characteristics diverged.)
Moving to the BMA BMDL, however, overcomes these negative characteristics: the BMA BMDL boxplot at far right is almost identical to that for the correct-model M$_3$ boxplot. Here again, the BMA BMDL provides valuable robustness in the presence of model uncertainty/misspecification.

\section{Discussion}\label{sec:disc}

\indent\indent Herein, we consider a strategy for model averaging within a hierarchical Bayesian framework when estimating benchmark doses (BMDs) in quantitative risk analysis. Placing emphasis on biomedical risk assessment, our approach estimates the BMD from a series of reparameterized quantal-response models with meaningful parameters (including the BMD itself) and accounts for potential model uncertainty via Bayesian model averaging (BMA).
A mixture posterior density is constructed for the BMD, and is used to find the BMA point estimate \citep{homa99}. The BMA lower credible limit (BMDL) is estimated as the lower $\alpha\mbox{th}$ percentile of the mixture posterior.  Risk analysts can apply this method to avoid concerns of model uncertainty in multimodel problems, and to construct inferences on the BMD by incorporating prior knowledge for the uncertainty associated with pertinent model parameters.

Of course, some caveats and qualifications are in order. Our reparameterized dose-response models are highly complex and for the three-parameter forms they can become rather unwieldy. For instance, \citet{fang14} notes that reparameterizations using $\gamma_1$ can produce unstable estimates when the highest dose level in the design is very close to the estimated BMD.  In a series of additional simulations (results not shown) we constructed such a parameter setting where the generating value of $\xi_{10}$ was slightly larger than $d_m$. We found that use of the three-parameter models often resulted in algorithm failures when an individual-model $\hat{\xi}$ drew near to $d_m$. Thus in cases where the BMD is felt to be near the highest dose---as determined, e.g., from elicited prior information---we recommend revising the reparameterization for $\gamma_1$ to define it as the response at some lower dose, say, $R(d_2)$. This can ameliorate the instabilities that occur when $\hat{\xi}$ is near $d_m$.

It is also of interest to investigate how our approach operates under different design configurations. We focused on a geometric, four-dose design, arguably the quintessential standard in cancer and laboratory-animal toxicity testing. We may gain greater information about the pattern of dose response and therefore about the BMD, however, if we increase the number of doses and/or change the dose spacings. Experimental design for dose-response studies with focus on the BMD is an emerging area in the statistical literature \citep{muri09,ober10,sand08,shsm12} and how to optimally design/allocate experimental resources for BMD estimation and inferences under a Bayesian paradigm is an emerging question.

\section*{\large Acknowledgements}

\begin{small}
Thanks are due Drs.~Katherine Y.~Barnes, Anton Westveld and D.~Dean Billheimer for their helpful suggestions during the preparation of this material.  The results represent a portion of the first author's Ph.D.~dissertation with the University of Arizona Graduate Interdisciplinary Program in Statistics.
\end{small}

\nocite*
\renewcommand\bibfont{\small}


\begin{thebibliography}{}
\begin{small}
\bibitem[\protect\citename{Akaike, }1973]{akai73} Akaike, H. (1973). \newblock{ Information theory and an extension of the maximum likelihood principle.} In \newblock{\em Proceedings of the Second International Symposium on Information Theory} (Petrov B.~N. and Csaki B., eds), 267--281. \newblock{ Akademiai Kiado, Budapest}.

\bibitem[\protect\citename{Andrieu and Thoms, }2008]{anth08} Andrieu, C. and Thoms, J. (2008). \newblock{ A tutorial on adaptive MCMC.} \newblock{{\em Statistics and Computing}} {\bf 18}, 343--383.

\bibitem[\protect\citename{Armitage and Doll, }1954]{ardo54} Armitage, P. and Doll, R. (1954). \newblock{ The age distribution of cancer and a multi-stage theory of carcinogenesis.} \newblock{{\em British Journal of Cancer}} {\bf 8}, 1--12.


\bibitem[\protect\citename{Bailer {\em et~al.}, }2005]{bano05} Bailer, A.~J., Noble, R.~B. and Wheeler, M.~W. (2005). \newblock{ Model uncertainty and risk estimation for experimental studies of quantal responses.} \textit{Risk Analysis} \textbf{25}, 291--299.








\bibitem[\protect\citename{Casella and Berger, }2002]{cabe02} Casella, G. and Berger, R.~L. (2002).  \newblock {\em Statistical Inference}, 2nd edn. \newblock Pacific Grove, CA: Duxbury.


\bibitem[\protect\citename{Christensen {\em et~al.}, }2011]{chjo11} Christensen, R., Johnson, W.~O., Branscum, A.~J. and Hanson, T.~E. (2011). \newblock {\em Bayesian Ideas and Data Analysis: An Introduction for Scientists and Statisticians}. \newblock Boca Raton, FL: Chapman \& Hall/CRC Press.



\bibitem[\protect\citename{Crump, }1984]{crum84} Crump, K.~S. (1984). \newblock{ A new method for determining allowable daily intake.} \newblock {\em Fundamental and Applied Toxicology} {\bf 4}, 854--871.

\bibitem[\protect\citename{Crump, }1995]{crum95} Crump, K.~S. (1995). \newblock{ Calculation of benchmark doses from continuous data.} \newblock {\em Risk Analysis} {\bf 15}, 79--89.



\bibitem[\protect\citename{Davis {\em et~al.}, }2012]{davi12} Davis, J.~A., Gift, J.~S. and Zhao, Q.~J. (2012). \newblock{ Introduction to benchmark dose methods and U.S. EPA's Benchmark Dose Software (BMDS) version 2.1.1.} \newblock {\em Toxicology and Applied Pharmacology} {\bf 254}, 181--191.

\bibitem[\protect\citename{European Union, }2003]{eu03} European Union (2003). \textit{ Technical Guidance Document (TGD) on Risk Assessment of Chemical Substances following European Regulations and Directives, Parts I-IV}. \newblock{ Technical Report number EUR 20418 EN/1-4.} \newblock Ispra, Italy: European Chemicals Bureau (ECB).


\bibitem[\protect\citename{Fang, }2014]{fang14} Fang, Q. (2014). \textit{Hierarchical Bayesian Benchmark Risk Analysis}. \newblock {Ph.D. thesis, Interdisciplinary Program in Statistics, University of Arizona, Tucson, AZ.}

\bibitem[\protect\citename{Fang and Piegorsch, }2014]{fapi14} Fang, Q. and Piegorsch, W.~W. (2014). \newblock{ Bayesian Benchmark Dose Analysis}. Submitted.

\bibitem[\protect\citename{Faustman and Bartell, }1997]{faus97} Faustman, E.~M. and Bartell, S.~M. (1997). \newblock{ Review of noncancer risk assessment: Applications of benchmark dose methods.} \newblock {\em Human and Ecological Risk Assessment} {\bf 3}, 893--920.








\bibitem[\protect\citename{Geweke, }1992]{gewe92} Geweke, J. (1992). \newblock{Evaluating the accuracy of sampling-based approaches to the calculation of posterior moments.} \newblock{In \textit{Bayesian Statistics} \textbf{4} (Bernardo, J.~M., Berger, J.~O., Dawid, A.~P. and Smith, A.~F.~M., eds.), 169--193.} \newblock{Oxford University Press, Oxford.}






\bibitem[\protect\citename{Guha {\em et~al.}, }2013]{guha13} Guha, N., Roy, A.,  Kopylev, L., Fox, J. Spassova, M. and White P. (2013). \newblock{Nonparametric Bayesian Methods for Benchmark Dose Estimation.} \newblock{\em Risk Analysis} {\bf 33}, 1608--1619.





\bibitem[\protect\citename{Hoeting {\em et~al.}, }1999]{homa99} Hoeting, J.~A., Madigan, D., Raftery, A.~E. and Volinsky, C.~T. (1999). \newblock{ Bayesian model averaging: A tutorial.} \textit{Statistical Science} \textbf{14}, 382--401. (corr. \textbf{15}, 193--195).





\bibitem[\protect\citename{Kang {\em et~al.}, }2000]{kako00} Kang, S.~H., Kodell. R.~L. and Chen, J.~J. (2000). \newblock{ Incorporating model uncertainties along with data uncertainties in microbial risk assessment.} \newblock{\em Regulatory Toxicology and Pharmacology} \textbf{32}, 68--72.






\bibitem[\protect\citename{Lambert {\em et~al.}, }2005]{lamb05} Lambert, P.~C., Sutton, A.~J., Burton, P.~R., Abrams, K.~R. and Jones, D.~R. (2005). \newblock{ How vague is vague? A simulation study of the impact of the use of vague prior distributions in MCMC using WinBUGS.} \newblock {\em Statistics in Medicine} {\bf 24}, 2401--2428.



\bibitem[\protect\citename{Lopes and West, }2004]{lope04} Lopes, H.~F. and West, M. (2004). \newblock{ Bayesian model assessment in factor analysis.} \newblock{\em Statistica Sinica} {\bf 14}, 41--67.

\bibitem[\protect\citename{Meng and Wong, }1996]{meng96} Meng, X. and Wong W.~H. (1996). \newblock{ Simulating ratios of normalizing constants via a simple identity: A theoretical exploration.} \newblock{\em Statistica Sinica} {\bf 6}, 831--860.



\bibitem[\protect\citename{Morales {\em et~al.}, }2006]{moib06} Morales, K.~H., Ibrahim, J.~G., Chen, C.-J. and Ryan, L.~M. (2006).  \newblock{ Bayesian model averaging with applications to benchmark dose estimation for arsenic in drinking water.} \newblock {\em Journal of the American Statistical Association} {\bf 101}, 9--17.

\bibitem[\protect\citename{Muri {\em et~al.}, }2009]{muri09} Muri, S.~D., Schlatter, J.~R. and Br$\ddot{\mbox{u}}$schweiler, B.~J. (2009). \newblock{ The benchmark dose approach in food risk assessment:  Is it applicable and worthwhile?} \newblock{\em Food and Chemical Toxicology} {\bf 47}, 2906--2925.


\bibitem[\protect\citename{Naufal {\em et~al.}, }2009]{nauf09} Naufal, Z., Kathman, S. and Wilson, C. (2009). \newblock{ Bayesian derivation of an oral cancer slope factor distribution for 4-(methylnitrosamino)-1-(3-pyridyl)-1-butanone (NNK).} \newblock {\em Regulatory Toxicology and Pharmacology} {\bf 55}, 69--75.

\bibitem[\protect\citename{Nitcheva {\em et~al.}, }2007]{nitc07} Nitcheva, D.~K., Piegorsch, W.~W. and West, R.~W. (2007). \newblock{ On use of the multistage dose-response model for assessing laboratory animal carcinogenicity.} \newblock {\em Regulatory Toxicology and Pharmacology} {\bf 48}, 135--147.

\bibitem[\protect\citename{\"{O}berg, }2010]{ober10} \"{O}berg M. (2010). \newblock{ Benchmark dose approaches in chemical health risk assessment in relation to number and distress of laboratory animals.} \newblock{\em Regulatory Toxicology and Pharmacology} {\bf 58}, 451--454.

\bibitem[\protect\citename{OECD, }2006]{oecd06} OECD (2006). \textit{ Current Approaches in the Statistical Analysis of Ecotoxicity Data: A Guidance to Application, Series on Testing and Assessment No. 54}. \newblock Paris: Environment Directorate, Organisation For Economic Co-Operation and Development.

\bibitem[\protect\citename{OECD, }2008]{oecd08} OECD (2008). \textit{ Draft Guidance Document on the Performance of Chronic Toxicity and Carcinogenicity Studies, Supporting TG 451, 452 and 453}. \newblock Paris: Organisation For Economic Co-Operation and Development.

\bibitem[\protect\citename{O'Hagan, }1994]{ohag94} O'Hagan, A. (1994). \textit{Kendall's Advanced Theory of Statistics, Volume 2B, Bayesian Inference}, 2nd edn. London: Edward Arnold.




\bibitem[\protect\citename{Piegorsch and Bailer, }2005]{piba05} Piegorsch, W.~W. and Bailer, A.~J. (2005). \newblock {\em Analyzing Environmental Data}. \newblock Chichester: John Wiley \& Sons.


\bibitem[\protect\citename{Piegorsch {\em et~al.}, }2013]{pian13} Piegorsch, W.~W., An, L., Wickens, A.~A., West, R. W. Pe$\tilde{\mbox{n}}$a, E. A. and Wu, W. (2013). \newblock{ Information-theoretic model-averaged benchmark dose analysis in environmental risk assessment}. \newblock{\em Environmetrics}, {\bf 24}, 143--157.

\bibitem[\protect\citename{Portier, }1994]{port94} Portier, C.~J. (1994). \newblock{ Biostatistical issues in the design and analysis of animal carcinogenicity experiments}. \newblock{\em Environmental Health Perspectives} {\bf102}, Suppl.~1, 5--8.

\bibitem[\protect\citename{{R Development Core Team}, }2012]{r12} {R Development Core Team}. (2012). \newblock {\em R: A Language and Environment for Statistical Computing}. \newblock Vienna, Austria: R Foundation for Statistical Computing. \newblock {ISBN} 3-900051-07-0.


\bibitem[\protect\citename{Robert and Casella, }2011]{roca11} Robert, C.~P. and Casella, G. (2011). \newblock{ A history of Markov chain Monte Carlo: subjective recollections from incomplete data.} \textit{Statistical Science} \textbf{26}, 102--115.


\bibitem[\protect\citename{Sand {\em et~al.}, }2008]{sand08} Sand, S., Victorin, K. and Falk~Filipsson, A. (2008). \newblock{ The current state of knowledge on the use of the benchmark dose concept in risk assessment.} \newblock{\em Journal of Applied Toxicology} \textbf{28}, 405--421.

\bibitem[\protect\citename{Shao, }2012]{shao12} Shao, K. (2012). \newblock{  A comparison of three methods for integrating historical information for Bayesian model averaged benchmark dose estimation.}  \textit{Environmental Toxicology and Pharmacology} \textbf{34}, 288--296.

\bibitem[\protect\citename{Shao and Gift, }2014]{shgi14} Shao, K. and Gift, J. S. (2014). \newblock{Model uncertainty and Bayesian model averaged benchmark dose estimation for continuous data.}  \textit{Risk Analysis} \textbf{34}, 101--120.

\bibitem[\protect\citename{Shao and Small, }2011]{shsm11} Shao, K. and Small, M.~J. (2011). \newblock{ Potential uncertainty reduction in model-averaged benchmark dose estimates informed by an additional dose study.} \newblock {\em Risk Analysis} {\bf 31}, 1561--1575.

\bibitem[\protect\citename{Shao and Small, }2012]{shsm12} Shao, K. and Small, M.~J. (2012). \newblock{ Statistical evaluation of toxicological experimental design for Bayesian model averaged benchmark dose estimation with dichotomous data.} \newblock {\em Human and Ecological Risk Assessment} {\bf 18}, 1096--1119.



\bibitem[\protect\citename{Stern, }2008]{ster08} Stern, A.~H. (2008). \newblock{ Environmental health risk assessment.} \newblock {In {\em Encyclopedia of Quantitative Risk Analysis and Assessment} {\bf 2} (Melnick, E.~L. and Everitt, B.~S., eds.), 580--589.} \newblock{John Wiley \& Sons, Chichester.}


\bibitem[\protect\citename{U.S.~EPA, }2012]{epa12} U.S.~EPA (2012). \newblock{ \em Benchmark Dose Technical Guidance Document}. \newblock{ Technical Report number EPA/100/R-12/001.} \newblock Washington, DC: U.S. Environmental Protection Agency.


\bibitem[\protect\citename{U.S.~General Accounting Office, }2001]{gao01} U.S.~General Accounting Office (2001). \textit{ Chemical Risk Assessment.  Selected Federal Agencies' Procedures, Assumptions, and Policies}. \newblock{ Report to Congressional Requesters number GAO-01-810.} \newblock Washington, DC: U.S.~General Accounting Office.

\bibitem[\protect\citename{U.S.~NTP, }2009]{tr542} U.S.~National Toxicology Program (2009). \newblock{ \em Toxicology and Carcinogenesis Studies of Cumene (CAS NO. 98-82-8) in F344/N Rats and B6C3F$_1$ Mice}. \newblock{ Technical Report number 542.} \newblock Research Triangle Park, NC: U.S. Department of Health and Human Services, Public Health Service.








\bibitem[\protect\citename{West {\em et~al.}, }2012]{wepi12} West, R.~W., Pigorsch, W.~W., Pe$\tilde{\mbox{n}}$a, E. A., An, L., Wu, W., Wickens, A.~A., Xiong, H., Chen, W. (2012). \newblock{ The impact of model uncertainty on benchmark dose estimation.} \newblock {\em Environmetrics} {\bf 23}, 706--716.



\bibitem[\protect\citename{Wheeler and Bailer, }2009]{whba09b} Wheeler, M.~W. and Bailer, A.~J. (2009). \newblock{ Benchmark dose estimation incorporating multiple data sources.} \newblock{\em Risk Analysis} {\bf 29}, 249--256.

\bibitem[\protect\citename{Wheeler and Bailer, }2012]{whba12} Wheeler, M.~W. and Bailer, A.~J. (2012). \newblock{ Monotonic Bayesian semiparametric benchmark dose analysis.} \newblock{\em Risk Analysis} {\bf 32}, 1207--1218.


\end{small}
\end{thebibliography}
\end{document}


\noindent \textbf{\large{Supplement to `Bayesian Model-Averaged Benchmark Dose Analysis Via Reparameterized Quantal-Response Model'}}

\vspace{8 pt}
\renewcommand\thefootnote{\fnsymbol{footnote}}
\noindent \textbf{Qijun Fang}$^{\textrm{a}}$, \textbf{Walter W. Piegorsch}$^{\textrm{a,b}}$, \textbf{Susan J. Simmons}$^{\textrm{c}}$, \textbf{Xiaosong Li}$^{\textrm{c}}$, \textbf{Cuixian Chen}$^{\textrm{c}}$ and \textbf{Yishi Wang}$^{\textrm{c}}$
\\
\noindent $^{\textrm{a}}$\textit{Program in Statistics} and $^{\textrm{b}}$\textit{BIO5 Institute}\\
\textit{University of Arizona, Tucson, AZ, USA}\\
\noindent $^{\textrm{c}}$\textit{Department of Mathematics and Statistics}\\
\textit{University of North Carolina at Wilmington, Wilmington, NC, USA}\\

\normalsize
\noindent This supplementary document provides supporting material for \emph{Bayesian Model-Averaged Benchmark Dose Analysis Via Reparameterized Quantal-Response Models} by Qijun Fang et.~al., (the `main document').  The various sections below address a variety of supplemental/supporting topics, and are not intended to flow naturally between each other.  They are, however, presented in roughly the same order in which their counterpart topics appeared in the main document.  As there, we denote the Benchmark Dose (BMD) target parameter as $\xi$ at a Benchmark Response (BMR) between 0 and 1, the background response probability nuisance parameter as $\gamma_0 = R(0)$ and the response at the highest dose, $d_m$, as $\gamma_1 = R(d_m)$.

\section{Example of reparameterizing Model M$_5$, the two-stage model}\label{sec:suppl1}

Here we study the two-stage model (M$_5$) to illustrate the reparameterization process, originally proposed in \citet{fapi14}. The risk function is
\begin{equation}\label{oldmulti}
R(d)=1-\exp(-\beta_0-\beta_1d-\beta_2d^2),
\end{equation}
where $\beta_j \ge 0 \;(j=0,1,2)$. With this, $\gamma_0 = R(0) = 1-\exp(-\beta_0)$, hence $\exp(-\beta_0)=1-\gamma_0$. Substitute $\exp(-\beta_0)=1-\gamma_0$ into the risk function to find
\begin{equation}\label{newmulti}
R(d)=\gamma_0+(1-\gamma_0)[1-\exp\{-\beta_1d-\beta_2d^2\}].
\end{equation}
Next, we reexpress $\beta_1$ and $\beta_2$ in terms of $\xi$, $\gamma_0$ and $\gamma_1$. Begin with $\gamma_1$, expressed in terms of $\gamma_0$, $\beta_1$ and $\beta_2$. From the definition of $\gamma_1$, we know
\[\gamma_1=\gamma_0+(1-\gamma_0)[1-\exp\{-\beta_1d_m-\beta_2d_m^2\}],\]
where $d_m$ denotes the $m$th (highest) dose level. Let $\Gamma_5 = \log\left(\frac{1-\gamma_1}{1-\gamma_0}\right)$. After some algebra, we obtain
\begin{equation}\label{GM5}
\beta_2d_m^2+\beta_1d_m+\Gamma_5=0.
\end{equation}
Now, \citet[Ex. 4.12]{piba05} show that the BMD for the two-stage model can be found to be
\begin{equation}\label{bmd5}
\xi=\frac{-\beta_1+\sqrt{\beta_1^2-4\beta_2\ln(1-\mbox{BMR})}}{2\beta_2}.
\end{equation}
For simplicity, denote  $C_5=-\ln(1-\mbox{BMR})$. Then from Equation (\ref{bmd5}), we have
\begin{equation}\label{xi5}
\beta_2\xi^2+\beta_1\xi-C_5=0 .
\end{equation}
If we solve Equation (\ref{GM5}) for $\beta_1$, we find $\beta_1=(-\Gamma_5-\beta_2d_m^{\,2})/d_m$. Substitute the result into \eqref{xi5} to achieve
\begin{equation}\label{beta25}
\beta_2 = \frac{\Gamma_5\xi+C_5 d_m}{\xi d_m(\xi-d_m)}.
\end{equation}
Similarly, if we solve Equation (\ref{GM5}) for $\beta_2$, then $\beta_2 = (-\Gamma_5-\beta_1d_m)/d_m^{\,2}$. Substitute the result into \eqref{xi5} to achieve
\begin{equation}\label{beta15}
\beta_1=\frac{C_5d_m^2+\Gamma_5\xi^2}{\xi d_m(d_m-\xi)}.
\end{equation}

So far, we have rewritten $\beta_1$ and $\beta_2$ in terms of $\Gamma_5$ (or $\gamma_0$ and $\gamma_1$) and $\xi$. The reparameterized risk function is obtained by substituting these expressions into Equation (\ref{newmulti}).  This produces
\begin{equation} \label{risk5}
R(d) = \gamma_0+(1-\gamma_0)\left[1-\exp\left\{\frac{C_5d_m d (d_m-d)+\Gamma_5 \xi d (\xi-d)}{\xi d_m(\xi-d_m)}\right\}\right],
\end{equation}
where $\Gamma_5=\log\left(\frac{1-\gamma_1}{1-\gamma_0}\right)$ and $C_5=-\log(1-\mbox{BMR})$.  Similar sorts of operations may be applied to reparameterize any of the quantal-response models in main document Table 2 \citep{fang14}.

\section{Finding prior parameters from elicited quantiles}\label{sec:suppl2}

\indent \indent We give here technical aspects on derivation of the prior parameters $\alpha$, $\beta$, $\psi$, $\omega$, $\kappa$ and $\lambda$ for the prior densities $\xi \sim \mbox{\textit{IG}}(\alpha, \beta)$, $\gamma_0 \sim \mbox{\textit{Beta}}(\psi, \omega)$ and $\gamma_1 \sim \mbox{\textit{Beta}}(\kappa, \lambda)$ used in the main document's hierarchical model.  The elicited lower quartiles and medians $Q_{1\xi}, Q_{2\xi}$ for $\xi$, $Q_{1\gamma_0}, Q_{2\gamma_0}$ for $\gamma_0$, and $Q_{1\gamma_1}, Q_{2\gamma_1}$ for $\gamma_1$, respectively, are assumed given from domain-expert input and/or previous literature/data.

Start with the elicitation for $\xi$:  by definition, $Q_{1\xi}$ and $Q_{2\xi}$ satisfy
\begin{equation*}\label{eq:xi.1}
\int_0^{Q_{j\xi}}\frac{\beta^\alpha}{\Gamma(\alpha)} \xi^{(-\alpha-1)} \exp\left\{-\frac{\beta}{\xi}\right\}d\xi = \frac{j}{4},
\end{equation*}
($j = 1,2$), where the integrand is the IG p.d.f.  This establishes a system of non-linear equations for $\alpha$ and $\beta$.  Unfortunately, no closed-form solution exists for the system, and so we turn to numerical methods.  We employ a gradient method discussed by \cite{barz88} and implemented in the \verb|R| statistical environment \citep{r12} via the \verb|BB| package \citep{vara09}. This method employs iterative updating until half the $L_2$ norm of the system at the proposed solution is smaller than $10^{-10}$.  The iterative solution represents the `elicited' values for $\alpha$ and $\beta$.

For $\gamma_0$, the elicited quartiles $Q_{1\gamma_0}$ and $Q_{2\gamma_0}$ satisfy
\begin{equation*}\label{eq:xi.2}
\int_0^{Q_{j\gamma_0}}\frac{\Gamma(\psi+\omega)}{\Gamma(\psi)\Gamma(\omega)} \gamma_0^{\psi-1}(1-\gamma_0)^{\omega-1}d\gamma_0 = \frac{j}{4},
\end{equation*}
($j = 1,2$), now applying the p.d.f.~for the Beta prior. Here again, the resulting system of non-linear equations possesses no closed-form solution, so we appeal to the \verb|BB| package in \verb|R| to produce the elicited values for $\phi$ and $\omega$.  In similar fashion, and where necessary, given elicited quartiles $Q_{1\gamma_1}$ and $Q_{2\gamma_1}$ for $\gamma_1$, we solve for $\kappa$ and $\lambda$ in $\gamma_1 \sim {Beta}(\kappa,\lambda)$.

The resulting iterative solutions to these systems of equations produce values for the prior parameters $\alpha$, $\beta$, $\psi$, $\omega$, $\kappa$ and $\lambda$ which are then employed in our hierarchical model for the BMD.

\section{Screen to check for 'data failures'}\label{sec:suppl3}

\indent \indent As described in \citet{fapi14}, some patterns of dose response can be problematic for benchmark dose estimation.  Our experience shows that a shallow dose response can create unstable frequentist $\hat{\xi}_{\mbox{\tiny 100BMR}}$ estimates, and very shallow responses may cause the iterative model fit to fail entirely.
Indeed, the EPA's BMDS software program will not estimate a BMD for a flat or negatively trending dose response \citep{whba09a}. For these reasons, we  mark flat or decreasing-trend data as `data failures' and not perform estimation for them.

To check the acceptability of the data before employing the algorithm, our simple screen is performed as follows: Over increasing doses $0=d_1<\ldots<d_i<\ldots<d_m$ $(i=1, \ldots m)$, we first calculate the empirical extra risks:
\[\tilde{R}_E(d_i)=\frac{\frac{Y_i}{N_i}-\frac{Y_1}{N_1}}{1-\frac{Y_1}{N_1}}.\]
We then connect each point $(d_i, \tilde{R}_E(d_i))$ to the origin point and find the largest slope, $S_{\scriptsize \mbox{max}}$, among all $m-1$ rays. If $S_{\scriptsize \mbox{max}}\le0$, no increasing trend is evidenced and we mark this as a `data failure.' Notice that we do not perform a formal trend test to detect a decreasing trend \citep{whba09a}, but we do require that at least one estimated risk at some $d_i$ ($i>1$) is higher than the estimated background risk.

\section{The adaptive Metropolis (AM) algorithm}\label{sec:suppl4}

\indent \indent For data passing the simple `data failure' screen in \S\ref{sec:suppl3}, we perform posterior simulation via a global AM procedure with componentwise adaptive scaling described by \citet{anth08}. Their `AM6' algorithm uses all previous iterations in the chain to update the variance-covariance matrix of the current bivariate proposal density.  The operation is global in the sense that all the parameters (here, $\xi$ and $\gamma_0$ for our two-parameter models and $\xi$, $\gamma_0$ and $\gamma_1$ for our three-parameter models) are updated simultaneously at each bivariate or trivariate draw.  It is componentwise in the sense that the updating step sizes for $\xi$ and $\gamma_0$ (or $\xi$, $\gamma_0$ and $\gamma_1$ for the three-parameter models) are controlled by separate adaptive scaling parameters. See \cite{anth08} for further details.

Denote ${\boldsymbol \theta}$ as the parameter vector ($[\xi~\gamma_0]^{{}^\text{\scriptsize T}}$ for the two-parameter case or $[\xi~\gamma_0\;\gamma_1]^{{}^\text{\scriptsize T}}$ for the three-parameter case).  Let $v_{k,1}, \ldots, v_{k,U}$ be adaptive scaling parameters and set $V_k=\mbox{diag}\left(v_{k,1}, \ldots, v_{k,U}\right)$ at the $k$th draw in the Monte Carlo chain ($k=1,\ldots,K$).  ($U$ denotes the dimension for the parameter vector; $U=2$ for our two-parameter models and $U=3$ for the three-parameter models.)  Take $\alpha$ as the acceptance probability in the normal random-walk Metropolis sampler, i.e.,
\[\alpha({\boldsymbol \theta}_k, {\boldsymbol \theta}_k+Z_{k+1}) = \min\left\{1, \frac{\pi({\boldsymbol \theta}_k|{\boldsymbol Y})}{\pi({\boldsymbol \theta}_k+Z_{k+1}|{\boldsymbol Y})}\right\}.\]
For notation in what follows, set $\boldsymbol{e}_U$ as a length-$U$ canonical vector with $u$th element equal to 1 and all others 0.  Denote $s_{k}$ as a step-size parameter that makes $\{s_{k}\}_{k=1}^\infty$ a decaying sequence; for example, the deterministic specification $s_{k}=1/k^p$, with $p>0$ is employed here. [\cite{anth08} recommend setting $p>1/2$, while \cite{viho12} suggests $p=2/3$. Here, \citeauthor{viho12}'s suggestion is followed and $s_{k}=k^{-2/3}$ is used.]  A brief description of the AM algorithm follows.

\begin{itemize}
\item Find starting values for $\xi$, $\gamma_0$, and (where needed) $\gamma_1$ in the chain as follows:  take the starting point for $\gamma_0$ by shrinking $\frac{Y_1}{N_1}$ away from 0 or 1
    via $\frac{Y_1+0.25}{N_1+0.5}$. Similarly, take the starting point for $\gamma_1$ as $\frac{Y_j+0.25}{N_j+0.5}$, where the index $j$ indicates the $j^{\scriptsize \mbox{th}}$ data observation at which $S_{\scriptsize \mbox{max}}$ is attained (see the `simple screen' procedure described in \S\ref{sec:suppl3}). For the two-parameter models, the starting point for $\xi$ is taken as BMR/$S_{\scriptsize \mbox{max}}$. This is obtained by first treating $R_E(d)=S_{\scriptsize \mbox{max}}d$ as the approximate extra risk function and then solving for $d$ in the equation $\mbox{BMR}=S_{\scriptsize \mbox{max}}d$. For the three-parameter models, the starting point for $\xi$ is taken as BMR$/\left(\frac{\gamma_{10}-\gamma_{00}}{1-\gamma_{00}}\right)$ where $\gamma_{00}$ and $\gamma_{10}$ are the starting values for $\gamma_0$ and $\gamma_1$, respectively. These starting points collectively become the starting vector for the chain.  Denote this as ${\boldsymbol \theta_1}$ in the algorithm and let ${\mu_1=\boldsymbol \theta_1}$.

\item Following \cite{anth08}, let $\Sigma_1=\mbox{\bf I}$ where ${\bf I}$ is the $U\times U$ identity matrix. Next, set $v_{1,1} = \ldots = v_{1,U} = \frac{2.38^2}{U}$ as the initial scaling parameters. Let $V_1=\mbox{diag}\left(v_{1,1}, \ldots, v_{1,U}\right)$, so that $V_1^{1/2}\mbox{\bf I}\Sigma_1^{1/2}$ is the initial variance-covariance matrix for use in the joint proposal distribution.

\item Repeat for $k=2, \ldots, K$:
\begin{enumerate}
\item Given $\mu_{k-1}$, $\Sigma_{k-1}$ and $v_{k-1,1}, \ldots, v_{k-1,U}$, sample from a multivariate normal proposal distribution: $Z_{k}\sim N(0, V_{k-1}^{1/2}\Sigma_{k-1}V_{k-1}^{1/2})$ and set ${\boldsymbol \theta}_{k}={\boldsymbol \theta}_{k-1}+Z_{k}$ with probability $\alpha({\boldsymbol \theta}_{k-1}, {\boldsymbol \theta}_{k-1}+Z_{k})$. Otherwise, set ${\boldsymbol \theta}_{k}={\boldsymbol \theta}_{k-1}$.
\item Update as follows for $u=1, \ldots, U$:
\[\log(v_{k,u})=\log(v_{k-1,u})+s_{k}[\alpha({\boldsymbol \theta}_{k-1}, {\boldsymbol \theta}_{k-1}+Z_{k}(u)e_u)-\bar{\alpha}_{**}],\]
where $\bar{\alpha}_{**}$ is the acceptance rate of the candidate points generated at each iteration. \citet{anth08} suggest that $\bar{\alpha}_{**}=0.44$ provides good performance; we followed their suggestion here.
\item Update as follows for $\mu_{k}$ and $\Sigma_{k}$:
\[\mu_{k}=\mu_{k-1}+s_{k}({\boldsymbol \theta}_{k}-\mu_{k-1}),\]
\[\Sigma_{k}=\Sigma_{k-1}+s_{k}[({\boldsymbol \theta}_{k}-\mu_{k-1})({\boldsymbol \theta}_{k}-\mu_{k-1})^T-\Sigma_{k-1}]\]
\end{enumerate}
\end{itemize}

\section{Monte Carlo `burn-in' diagnostics}\label{sec:suppl5}

\indent \indent To account for potential instability in the early portions of the length-$K$, bivariate, AM chain from \S\ref{sec:suppl4}, we include a `burn-in' for the Monte Carlo sample \citep[\S11.6]{geca04}.  We couple this with the larger question of how to assess the chain's convergence.
Many approaches exist for diagnosing MCMC convergence \citep{coca96}, and indeed, the issue is a topic of ongoing debate \citep[\S12.2]{roca04}.
We mimic a method due to \citet{gewe92}, where early portions of the chain are sub-sampled and compared against latter portions to determine where the larger sample of $K$ draws begins to exhibit stable performance.  To approximate independence between the two sub-samples, we bifurcate the chain by a gap of no less than $\frac{K}{5}$ consecutive draws.

For summary diagnostics, we calculate the difference in arithmetic means between the two bifurcated sub-samples, and divide each difference by its standard error to produce a $Z$-statistic.  The approximate standard error is taken as the square root of the sum of the variances of each mean.  This is done separately for the $\xi_k$, $\gamma_{0k}$, and $\gamma_{1k}$  components.  To adjust for possible autocorrelation within each sub-sample, the individual variances are based on estimated spectral densities at frequency zero. Two parametric methods of estimating these spectral densities are considered here. The first method is discussed by \cite{hewe81}, which consists of fitting a generalized linear model to the periodogram of the batched time series.  We implement this using \verb|R|, via the \verb|coda| package's \verb|spectrum0| function. The second method begins by fitting the periodogram with an autogressive model, and then use the spectral density of the fitted AR model to approximate the spectral density at frequency zero.  We again implement this method using \verb|R|, now with the \verb|coda| package's \verb|spectrum0.ar| function. These two methods produced similar estimates of the spectral density at zero. The \citeauthor{hewe81} method is fast and is the default method in the \verb|coda| package, hence we also default to it first. The fit may fail due to high autocorrelation of the chain, however, and when this happens we then switch to the AR method for estimating the spectral density at zero.

To monitor if the pattern of association between any two parameters within the larger chain also exhibits reasonable stability, we include comparison of up to three pairs of covariances between two sub-samples (for the two-parameter models, only the covariance between $\xi$ and $\gamma_0$ is considered).  As a diagnostic measure we use the sample covariances from each sub-sample,  which are
\[\widehat{\mbox{Cov}}(\xi, \gamma_0)=\frac{1}{L}\sum_{k=1}^L(\xi_k-\bar{\xi})(\gamma_{0k}-\bar{\gamma}_0), \]
\[\widehat{\mbox{Cov}}(\xi, \gamma_1)=\frac{1}{L}\sum_{k=1}^L(\xi_k-\bar{\xi})(\gamma_{1k}-\bar{\gamma}_1), \]
\[\widehat{\mbox{Cov}}(\gamma_0, \gamma_1)=\frac{1}{L}\sum_{k=1}^L(\gamma_{0k}-\bar{\gamma}_0)(\gamma_{1k}-\bar{\gamma}_1), \]
where $\bar{\xi}$, $\bar{\gamma}_0$ and $\bar{\gamma}_1$ are the pertinent arithmetic means within a bifurcated sub-sample of length $L$ (see below).
To find the approximate variance of each estimated covariance we calculate the quantities
\[\tau_k(\xi, \gamma_0) = (\xi_k-\bar{\xi})({\gamma_{0k}}-\bar{\gamma}_0),\]
\[\tau_k(\xi, \gamma_1) = (\xi_k-\bar{\xi})({\gamma_{0k}}-\bar{\gamma}_1),\]
\[\tau_k(\gamma_0, \gamma_1) = (\gamma_{0k}-\bar{\gamma}_0)({\gamma_{1k}}-\bar{\gamma}_1),\]
across all values of the index $k$ in the given sub-sample and then estimate the variance of $\tau_k(\xi,\gamma_0)$, $\tau_k(\gamma_0,\gamma_1)$ and $\tau_k(\gamma_0,\gamma_1)$ based on their estimated spectral densities at frequency zero.

Each such comparison is conducted over three individual, separated bifurcations of the full $K$-length sample: (i) the first 10\% of the chain ($L = K/10$) vs.~the final 50\% ($L = K/2$), (ii) the first 20\% of the chain ($L = K/5$) vs.~the final 50\%, and (iii) the first 30\% of the chain ($L = 0.3K$) vs.~the final 50\%. We consider an individual diagnostic comparison as a `pass' if the corresponding $Z$-statistic is less than 1.96 in absolute value.  For the two-parameter models, to `pass' the full diagnostic at each bifurcation, all three measures---mean of $\xi$, mean of $\gamma_0$, and covariance of $\{\xi$,$\gamma_0\}$---must individually pass; for the three-parameter models, in addition, the mean of $\gamma_1$, and the covariances of $\{\xi, \gamma_1\}$ and $\{\gamma_0, \gamma_1\}$ must also pass.  The comparisons are performed in sequential order:  if the 10\%-vs.-50\% diagnostic fails, then the 20\%-vs.-50\% diagnostic is conducted. If this fails, we move to the 30\%-vs.-50\% diagnostic. When a diagnostic passes, we view the indicated early portion of the chain as a reasonable burn-in.  For notation, we let $K_0<K$ be the resulting index that begins the retained portion of the chain. Thus, e.g., if the 10\%-vs.-50\% diagnostic fails but then the 20\%-vs.-50\% diagnostic passes, we take the first 20\% of the chain as burn-in and use the remaining 80\% of the chain as our Monte Carlo sample from $\pi(\xi, \gamma_0|{\boldsymbol Y})$ or $\pi(\xi, \gamma_0, \gamma_1|{\boldsymbol Y})$. If $K=$100,000, as used throughout, this gives $K_0=$20,001.

If none of the three sequential diagnostics pass, we re-set the random-number generator's seed and reassess a new set of $K$ Monte Carlo draws.  If after five such re-starts the diagnostic continues to fail, we report an {\em algorithm failure}, and terminate the analysis with the given data set.

\section{Bridge sampling for estimating marginal distributions}\label{sec:suppl7}

\indent \indent As a necessary step to perform Bayesian model averaging, we must calculate an approximation to marginal likelihoods of the form (with the two-parameter models)
\[m(\boldsymbol Y) = \int_0^\infty\int_0^1 f({\boldsymbol Y}|\xi, \gamma_0)\pi(\xi|\alpha,\beta)\pi(\gamma_0|\psi,\omega)d\gamma_0d\xi\]
for elicited or objective priors $\pi(\xi|\alpha,\beta)$ and/or $\pi(\gamma_0|\psi,\omega)$.
To do so, we constructed a \emph{geometric estimator} from our AM sample, appealing to the `bridge sampling' method recommended by \cite{meng96} and \cite{lope04}.  To approximate the marginal likelihood, an approximation $g(\cdot)$ to the joint posterior density of $(\xi, \gamma_0)$ is first chosen. For simplicity, we took $g(\xi, \gamma_0)$ as the bivariate normal density function with mean set to the empirical mean vector and variance set to the empirical variance-covariance matrix of $\xi$ and $\gamma_0$, estimated from the AM sample. Next, a set of $K^* = K - K_0$ bivariate vectors $\big\{[\xi^*_j~ \gamma^*_{0j}]^{\mbox{\scriptsize \textit{T}}}\big\}_{j=1}^{K^*}$ are drawn from $g(\xi, \gamma_0)$. The marginal p.d.f.~is then approximated by substituting the (retained) AM sample, $\left\{\xi_k, \gamma_{0k}\right\}_{k=K_0}^K$, and the generated sample from $g(\xi, \gamma_0)$, $\big\{[\xi^*_j~ \gamma^*_{0j}]^{\mbox{\scriptsize \textit{T}}}\big\}_{j=1}^{K^*}$, into the ratio
$$\widehat{m}(\boldsymbol Y)=\frac{\frac{1}{K^*}\sum_{j=1}^{K^*}\left\{\pi\left(\xi^{*}_j|\alpha,\beta\right)\pi\left(\gamma_{0j}^{*}|\psi,\omega\right)f\left({\boldsymbol Y}|\xi^{*}_j,\gamma_{0j}^{*}\right)/g\left(\xi^{*}_j,\gamma_{0j}^{*}\right)\right\}^{1/2}}
{\frac{1}{K^*}\sum_{k=K_0}^K\left\{g\left(\xi_k,\gamma_{0k}\right)/\left[\pi\left(\xi_k|\alpha,\beta\right)\pi\left(\gamma_{0k}|\psi,\omega\right)f\left({\boldsymbol Y}|\xi_k,\gamma_{0k}\right)\right]\right\}^{1/2}},$$
where $f(\cdot)$ denotes the binomial likelihood function and $\pi(\cdot)$ denotes the pertinent prior density. The above expression gives the approximated marginal likelihood for our two-parameter models.  Marginal likelihoods for our three-parameter models are approximated in the same fashion by including the $\pi(\gamma_1|\kappa,\lambda)$ prior and using a trivariate approximating normal density $g(\xi, \gamma_0, \gamma_1)$.

\section{BMA BMDL calculation}\label{sec:suppl8}

\indent\indent In the main document, we claimed that
\[P(\xi\le \mbox{\underline{$\xi$}}_{\mbox{\tiny 100BMR}}|\boldsymbol{Y}, U_Q)\approx\sum_{q=1}^Q w_q\frac{1}{K^*_q}\sum_{j=1}^{K^*_q}I\{\xi_j\le\mbox{\underline{$\xi$}}_{\mbox{\tiny 100BMR}}|M_q\}.\]
Here we give a simple proof.
\begin{proof}
Equation (3.1) in the main document defines the model-averaged posterior density for the BMD, $f(\xi|\boldsymbol{Y}, U_Q)$, as a mixture of marginal posterior densities for $\xi$ under each $q$th model, $f(\xi|\boldsymbol{Y}, M_q)$ using each model's posterior probability $P(M_q|\boldsymbol{Y})$ as the corresponding weight, $w_q$.

Denote $F(\xi|\boldsymbol{Y}, U_Q)$ as the model-averaged posterior distribution function for the BMD and $F(\xi|\boldsymbol{Y}, M_q)$ as the marginal posterior distribution function for $\xi$ under each model.  By integrating both sides of Equation (3.1), we have
\begin{equation*}
F(\xi|\boldsymbol{Y}, U_Q)=\sum_{q=1}^Qw_qF(\xi|\boldsymbol{Y}, M_q).
\end{equation*}
Now, substitute \underline{$\xi$}$_{\mbox{\tiny 100BMR}}$ into the above equation to find
\begin{eqnarray*}
P(\xi\le \mbox{\underline{$\xi$}}_{\mbox{\tiny 100BMR}}|\boldsymbol{Y}, U_Q)&=&F(\mbox{\underline{$\xi$}}_{\mbox{\tiny 100BMR}}|\boldsymbol{Y}, U_Q)\\
&=&\sum_{q=1}^Qw_qF(\mbox{\underline{$\xi$}}_{\mbox{\tiny 100BMR}}|\boldsymbol{Y}, M_q)\\
&=&\sum_{q=1}^Qw_qP(\xi\le \mbox{\underline{$\xi$}}_{\mbox{\tiny 100BMR}}|\boldsymbol{Y}, M_q).
\end{eqnarray*}
Clearly then, the probability, $P(\xi\le \mbox{\underline{$\xi$}}_{\mbox{\tiny 100BMR}}|\boldsymbol{Y}, M_q)$ can be approximated by the sample proportion $\frac{1}{K^*_q}\sum_{j=1}^{K^*_q}I\{\xi_j\le\mbox{\underline{$\xi$}}_{\mbox{\tiny 100BMR}}|M_q\}$, where $K_q^*$ is the size of the AM sample under $M_q$ (excluding burn-in) and $I$ denotes the indication function.
\end{proof}

The BMA BMDL is then calculated by finding the root of equation
$$\frac{1}{K^*_q}\sum_{j=1}^{K^*_q}I\{\xi_j\le\mbox{\underline{$\xi$}}_{\mbox{\tiny 100BMR}}|M_q\}=0.05.$$
To solve this equation numerically, we use the \verb|R| \verb|uniroot| function. This employs iterative updating until either an exact root is found or the change in root for one step of the algorithm is less than $2^{\scriptsize -13}$.

\section{Prior elicitation with the cumene data}\label{sec:suppl9}

\indent \indent  To elicit the prior parameters for the cumene data in the main document, we applied input from domain experts and based the prior elicitation on existing background from the toxicological literature.
We began with the BMD target parameter, $\xi$:  an oral No Observed Adverse Effect Level (NOAEL) for long-term cumene exposure in rodents was given by the EPA IRIS website (\verb|http://www.epa.gov/iris/subst/0306.htm|) as 110 mg/kg-day. Guidance for converting  the oral mg/kg-day metric to inhalation ppm (as used in the original data Table from the main document) was available via conversion data given by the \cite{OEHHA04}:  for rodents, 1 mg/kg-day works out to roughly 2.25 ppm.  Arguing that a NOAEL is loosely equivalent to a BMD---although, see \citep{kode05} for an in-depth discussion---the median prior estimate was $Q_{2\xi} = 110$ mg/kg-day = 247.5 ppm $\approx 250$ ppm.  For the first quartile, the EPA IRIS site gave a cumene inhalation NOAEL for rodents as 435 mg/cu.m, based on a shorter, 13-week study. The shorter exposure period was suited to provide only a conservative estimate on the BMD's central tendency; we translated this to use of the 435 mg/cu.m 13-week NOAEL as a prior estimate of the first quartile.  For conversion to inhalation ppm, the EPA indicated that roughly 1 mg/cu.m = 0.204 ppm, so we took $Q_{1\xi} = 435$ mg/cu.m = 88.75 ppm $\approx 90$ ppm. We then divided the ppm dose by 500 ppm to standardize the $\xi$-scale to $0\le \xi \le1$. This produced prior first quartile and median estimates for $\pi(\xi|\alpha, \beta)$ as 0.18 and 0.5, respectively.

For $\gamma_0$, historical control data on alveolar/bronchiolar tumors in Female B6C3F$_1$ mice were given by
\citet[Table D3a]{tr542}. These provide direct prior information on the background risk.  The historical data gave a median background response of $Q_{2\gamma_0} = 0.08$ and a lower quartile of $Q_{1\gamma_0} = 0.04$.

As noted in the main document, no elicitation for $\gamma_1$ was available. We found no indication of carcinogenic effects of cumene evidenced in laboratory mice prior to the NTP's toxicology study \citep{tr542}. As a result, we defaulted to the objective prior $\gamma_1\sim Beta\left(\frac12, \frac12\right)$.

These quartile-based elicitations were then used to construct the prior densities $\xi \sim \mbox{\textit{IG}}(\alpha, \beta)$ and $\gamma_0 \sim \mbox{\textit{Beta}}(\psi, \omega)$ as described in \S\ref{sec:suppl1}, above.

\nocite*
\renewcommand\bibfont{\small}

\begin{footnotesize}

\end{footnotesize}